\documentclass{osa-article}
\newcommand*{\Journal}[1]{%
  \IfFileExists{#1.sty}{%
    \RequirePackage{#1}
  }{%
    \ClassError{osa-article}{You've specified an unsupported journal '#1'. Was that a typo, or are you using a copy of the template without the complete set of style files?}{Was that a typo, or are you using a copy of the template without the complete set of style files?}
  }
} 

\Journal{oe}

\usepackage[squaren]{SIunits}
\usepackage{hyphenat}
\usepackage[bookmarks=false]{hyperref}
\usepackage[capitalise]{cleveref}
\usepackage{graphicx}
\crefname{section}{Sec.}{Secs.}% APS style uses abbreviations
\Crefname{section}{Section}{Sections}
\usepackage{xcolor}
\usepackage{pifont}

\newcommand*\circled[1]{\raisebox{-1.65pt}{\Large\ding{\numexpr191+#1\relax}}}

\newcommand{\Chan}[1]{#1}

\begin{document}
\title{A countermeasure against bright-light attack on superconducting nanowire single-photon detector in quantum key distribution}

\author{Mikhail~Elezov,\authormark{1} Roman~Ozhegov,\authormark{1,2,*} Gregory~Goltsman,\authormark{1,2} and Vadim~Makarov\authormark{3,4,5,6}}

\address{\authormark{1}Moscow Pedagogical State University, Moscow 119991, Russia\\
\authormark{2}National Research University Higher School of Economics, Moscow Institute of Electronics and Mathematics, Moscow 109028, Russia\\
\authormark{3}Russian Quantum Center, Skolkovo, Moscow 143025, Russia\\
\authormark{4}Shanghai Branch, National Laboratory for Physical Sciences at Microscale and CAS Center for Excellence in Quantum Information, University of Science and Technology of China, Shanghai 201315, People's Republic of China\\
\authormark{5}NTI Center for Quantum Communications, National University of Science and Technology MISiS, Moscow 119049, Russia\\
\authormark{6}Department of Physics and Astronomy, University of Waterloo, Waterloo, ON, N2L~3G1 Canada}

\email{\authormark{*}ozhegov@rplab.ru}

\begin{abstract}
We present an active anti-latching system for superconducting nanowire single-photon detectors. We experimentally test it against a bright-light attack, previously used to compromise security of quantum key distribution. Although our system detects continuous blinding, the detector is shown to be partially blindable and controllable by specially tailored sequences of bright pulses. Improvements to the countermeasure are suggested.
\end{abstract}

\section{Introduction}

More than 30 years have passed since the method of transferring secret information based on the mechanisms of quantum physics, called quantum cryptography, was proposed \cite{bennett1984}. Since then, quantum cryptography has significantly developed: various protocols for transmitting key information have been created, various schemes for organizing quantum key distribution systems have been proposed, which have led to the creation of the first commercial quantum key distribution (QKD) systems \cite{kollmitzer2010}. At the same time, the development of QKD systems to the level of a commercial product has stimulated research into potential vulnerabilities of these systems to hacking attacks. It has been proved that QKD using perfect devices is secure  \cite{lo1999,shor2000}. Potential vulnerabilities depend on the technical realization of a specific QKD system and are exploited through its imperfection and operating aspects of the employed electronic and optical components. Searching and eliminating potential vulnerabilities in implementations are crucial for practical security. One of these vulnerable elements of a QKD system is a single-photon detector, of which the most commonly used types are single-photon avalanche diodes (SPADs) and superconducting nanowire single-photon detectors (SNSPDs). There are a number of attacks on QKD systems that use imperfections of the single-photon detector (for example, blinding, high-power damage, time-shift, after-gate attacks, superlinearity, efficiency mismatch and calibration loopholes \cite{jain2016}). 

A special place is occupied by the group of faked-state attacks on single-photon detectors in QKD systems \cite{makarov2005}. When implementing this type of attack, the eavesdropper Eve prepares optical pulses using a special algorithm and sends them to Bob (the nodes of the quantum key distribution system are usually referred to as Alice and Bob). With the successful implementation of this type of attack, the bases and states of the quantum states detected by Bob always coincide with the corresponding bases and states of the optical signals prepared by Eve, and the number of errors in Bob's key does not increase. After that, Eve can just listen to error correction and privacy amplification of the key made by Alice and Bob and can then apply the same operations to her key. As a result, Eve obtains a secret key identical to that of Alice and Bob. One of the options for implementing a faked-state attack is the blinding attack \cite{makarov2009,gerhardt2011,lydersen2010a}. A demonstration of hacking a QKD system using this approach was carried out in \cite{gerhardt2011}.

Apart from SPADs, SNSPDs are widely used in QKD systems. Compared to SPADs, the SNSPD can have a higher detection efficiency (above 90\%), a lower dark-count rate, a higher maximum count rate  and a lower jitter \cite{natarajan2012,marsili2013,smirnov2018,shcheslavskiy2016}. A blinding attack on the SNSPDs in a QKD system may be successful because of an imperfection of the SNSPD operation allowing it to be controlled by bright light \cite{lydersen2011c,tanner2014}. The potential vulnerabilities of the SNSPD are a latching effect and an ability to respond to bright optical pulses \cite{lydersen2011c}. The behavior of an SNSPD with a shunt resistor that prevents the latching effect has also been investigated \cite{tanner2014}. In that experiment, a light pattern that achieves temporary blinding without latching has been simulated, showing that a successful blinding attack on QKD may still be possible.

Here we discuss the development of a countermeasure against the blinding attack for a QKD system with SNSPDs. \Chan{The main goal of our work is to create a perfectly secure single-photon detector. Our testing is therefore focused on revealing any residual imperfections in the countermeasure, which if found shall lead to its redesign.} The countermeasure consists of hardware and software that form together an automatic anti-latching reset system in the bias supply of the SNSPD, presented in \cref{sec:setup}. We test this countermeasure against a bright-light attack in \cref{sec:experiment}. While the present implementation is found to allow a certain degree of controllability by Eve \Chan{and thus fails the test}, an improvement of the design is possible, as we discuss in \cref{sec:discussion}. We conclude in \cref{sec:conclusion}.

\section{Detector design and operation}
\label{sec:setup}

Our SNSPD is based on a superconducting current-carrying NbN film deposited on a polished Si wafer, with a superconducting transition critical temperature $T_{c} \approx 11~\kelvin$. To achieve the operating temperature we use a Gifford McMahon cryocooler with  a base temperature of $2.5~\kelvin$. The $4~\nano\meter$ thick NbN film is patterned into a $100~\nano\meter$ wide nanowire meander covering an area of $7 \times 7~\micro\meter\squared$ \cite{korneev2013}.

\begin{figure}
\centering
\includegraphics[width=86mm]{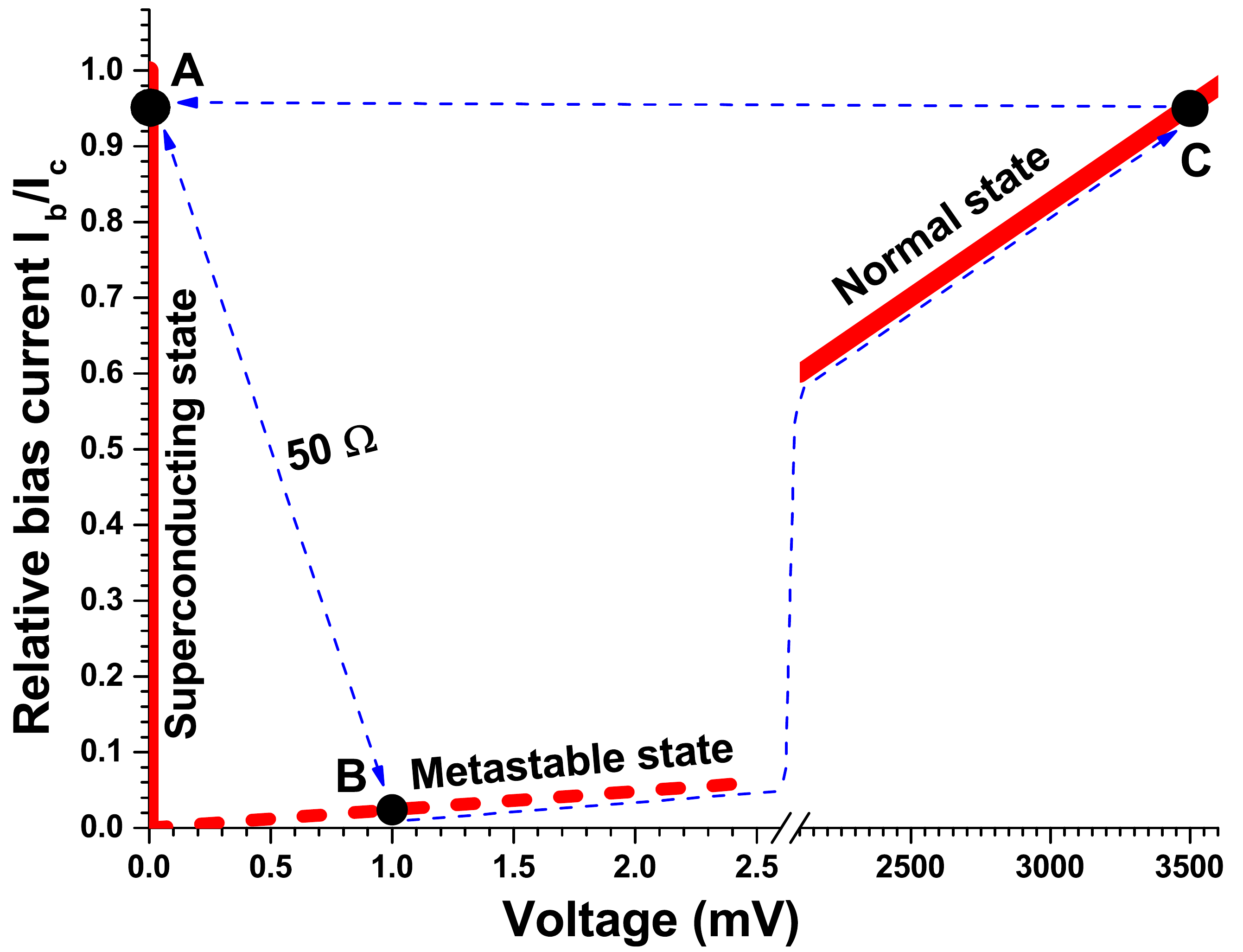}
\caption{\label{fig:IV_curve} 
A schematic current-voltage ($I$--$V$) characteristic of the SNSPD. Point~A is the SNSPD operating point. Point~B is a metastable state. Point~C is a latched state. The blue arrows show the evolution of the operating point towards the latched state.}
\end{figure}

\Cref{fig:IV_curve} shows a schematic current-voltage ($I$--$V$) curve of the SNSPD. The operating point of the detector (point~A) is typically at a bias current $I_b$ slightly below the critical current $I_c$. The typical value of $I_b = (0.90$--$0.95) I_c$. In this range, a high detection efficiency is combined with a low dark count rate. The SNSPD in our experiment has a system detection efficiency of $26.5\%$ and the dark count rate of $50~\hertz$ at $I_b = 22~\micro\ampere$.

When the bias current flows through the superconducting nanowire and the nanowire absorbs a photon, a local disturbance of the superconducting state in the nanowire appears. As a result, a normal domain is formed. Joule heating by the bias current makes the domain grow rapidly to the width of the nanowire \cite{semenov2001}. Because the normal-conducting nanowire has a high resistance, the bias current is reduced (point B). After that, superconductivity in the nanowire is recovered and the bias current returns to its original value (point A). The amplitude of the SNSPD response pulse is typically $1~\milli\volt$ (\cref{fig:IV_curve}). Its width is $\sim 7~\nano\second$ and depends on the kinetic inductance of the nanowire and thus on the size and shape of the detector.

After absorbing an optical pulse, the SNSPD may enter a latched state. The dissipation of the energy received by the detector from the optical pulse takes some time, which increases with the energy of the pulse. During the energy dissipation, the operating point of the detector drifts from point~B to point~C (with the normal domain continuing to grow in size). This leads to an increase in the power output and the further self-heating of the detector. This process happens with all types of bias sources; however, its parameters substantially depend on the type of the bias source used. A typical current source connected to a high impedance load as is the case here may develop several volts output voltage before it clamps. A stable state may thus arise in which the nanowire is blocked by the long normal domain and the superconducting state cannot be restored without externally reducing the bias current to zero. In this latched state, the SNSPD has low sensitivity and cannot detect single photons. However it can still operate in a bolometric mode, where its response may depend on the absorbed optical power. This property has been used to carry out the blinding attack \cite{lydersen2011c}. In order to avoid the latching effect, one can install a shunt resistor \cite{liu2012}, for example on the cable between the SNSPD and the bias tee \cite{tanner2014}. However, it has been shown that Eve can conduct a successful blinding attack even in this case, blinding the detector temporarily without achieving the latched state \cite{tanner2014}.

\Chan{To design a countermeasure against detector control, we have considered different approaches. We have ruled out those that involve adding components to the optical scheme, because any complication of the QKD system leads to the possible emergence of new attacks. One often suggested countermeasure of this kind is the addition of a beamsplitter and a monitoring detector that watches for abnormally strong input light \cite{lydersen2010a}. This countermeasure has extra drawbacks. Firstly, it leads to an additional loss in the signal path and thus lower QKD performance. Secondly, the sensitivity range of a classical photodetector is very different from that of the single-photon detector and may be insufficient to detect Eve's control signal. Should a single-photon detector be used as the monitoring detector, its security problems would apply. We have thus decided to leave the optical scheme intact, and monitor for a correct single-photon operation of the SNSPD by amending the electronics of its bias current supply.}

\begin{figure}
\centering
\includegraphics[width=90mm]{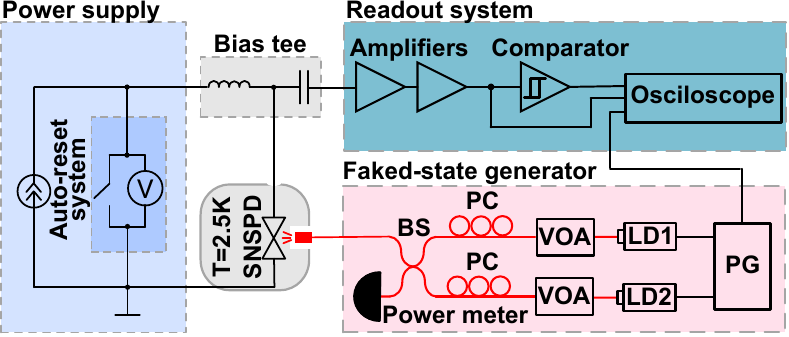}
\caption{\label{fig:Experimental_setup} 
Experimental setup. The SNSPD is cooled to $2.5~\kelvin$ and irradiated through single-mode fiber shown in red. The faked-state generator consists of two laser diodes (LD) producing linearly polarised $1550~\nano\meter$ light and controlled by a pulse generator (PG). The power and polarisation of each can be adjusted with variable optical attenuators (VOA) and polarisation controllers (PC). The laser power is divided by the $50\!:\!50$ beam splitter (BS) with one part fed into an optical power meter and the other sent to the SNSPD. The electronics consist of a power supply, a readout system and a bias tee with a bandwidth of $0.1$ to $4200~\mega\hertz$. The power supply consists of a highly stable and low-noise current supply and an auto-reset system, which includes an electronic switch and a fast voltmeter. The SNSPD response is amplified to $\sim 200~\milli\volt$ by two integrated amplifiers with a gain $\sim 40~\deci\bel$ and a bandwidth from $100~\mega\hertz$ to $3~\giga\hertz$, and is fed to a high-speed voltage comparator. We observed the amplified SNSPD response and the comparator output with an oscilloscope of $4~\giga\hertz$ bandwidth.}
\end{figure}

The power supply consists of a highly stable current source and our addition,  auto-reset system (\cref{fig:Experimental_setup}). The latter consists of a CMOS-based electronic switch (ADG619) and  a voltmeter. The voltmeter measures the voltage $V$ across the SNSPD through the bias tee (Mini-Circuits ZFBT-4R2GW+) and the coaxial cable leading into the cryostat. The setting of the current source and state of the switch are controlled by software running in a microcontroller.

The SNSPD response is amplified to $\sim 200~\milli\volt$ by two integrated amplifiers (Infineon BGA427) with a net  gain  of $\sim 40~\deci\bel$ and is fed to a high-speed voltage comparator (ADCMP601) with a threshold level of $60~\milli\volt$. In normal detector operation, a positive transition (low to high logic level) at the comparator output is regarded as a photon detection or `click'. We can observe the amplified SNSPD response and the comparator output with an oscilloscope (Rohde \& Schwarz RTO1044).

The auto-reset system allows us to avoid the SNSPD latching. The switch is normally open. The SNSPD voltage $V$ is measured with a sampling rate of $10~\kilo\hertz$, thereafter digitally low-pass filtered resulting in an effective sampling rate $\sim 100~\hertz$. If the SNSPD latches, $V$ becomes higher than an auto-reset trigger threshold (set at $20~\milli\volt$ in our experiment). \Chan{The above low-pass filtering and threshold voltage have been tuned to avoid the auto-reset triggering by voltmeter electronic noise and by voltage fluctuations produced by normal photon counting pulses.} When the auto-reset procedure is triggered, the switch is closed. The recovery time of the superconducting state of the nanowire depends on many parameters (for example on the power of the optical pulse $I_b$ and the topology of the detector). Therefore, we had an adjustable wait time in a range of $0.1$--$10~\second$ (in our case this time was set at $1~\second$). After this time the switch is opened, which completes the cycle. Any triggering of the auto-reset procedure should be time-stamped by a logging system that has not been included in the present experiment but since been implemented in a commercial SNSPD device \cite{scontel}. This would allow the QKD system to discard a portion of the raw key that may have been affected by the blinding attack. 

\section{Does the auto-reset system thwart the blinding attack?}
\label{sec:experiment}	

\subsection{Test setup}

In order to test the countermeasure, we have used a faked-state generator capable of forming pulse sequences of varying time and optical power, as well as continuous-wave light (\cref{fig:Experimental_setup}). We use two semiconductor laser diodes at $1550~\nano\meter$ (Gooch $\&$ Housego AA1406). A pulse generator (Highland Technology P400) controls the lasers, which emit rectangular optical pulses with a peak power $P_p$, width $\tau$, and pulse repetition rate $f$. The laser radiation is sent to a $50\!:\!50$ beam splitter. To measure $P_p$ of each laser, its integral power $P_i$ is measured by an optical power meter (Ophir VEGA with sensor PD300-IRG-V1). Then $P_p=P_i/(\tau f)$. We control $P_p$ of each laser by a variable optical fiber attenuator (EXFO FVA-600B). Since the SNSPD is sensitive to polarisation \cite{driessen2009}, a polarisation controller (Thorlabs FPC030) is used to ensure the maximum response of the detector to the incident light power.

\subsection{Latching threshold}

\begin{figure}
\centering
\includegraphics[width=78mm]{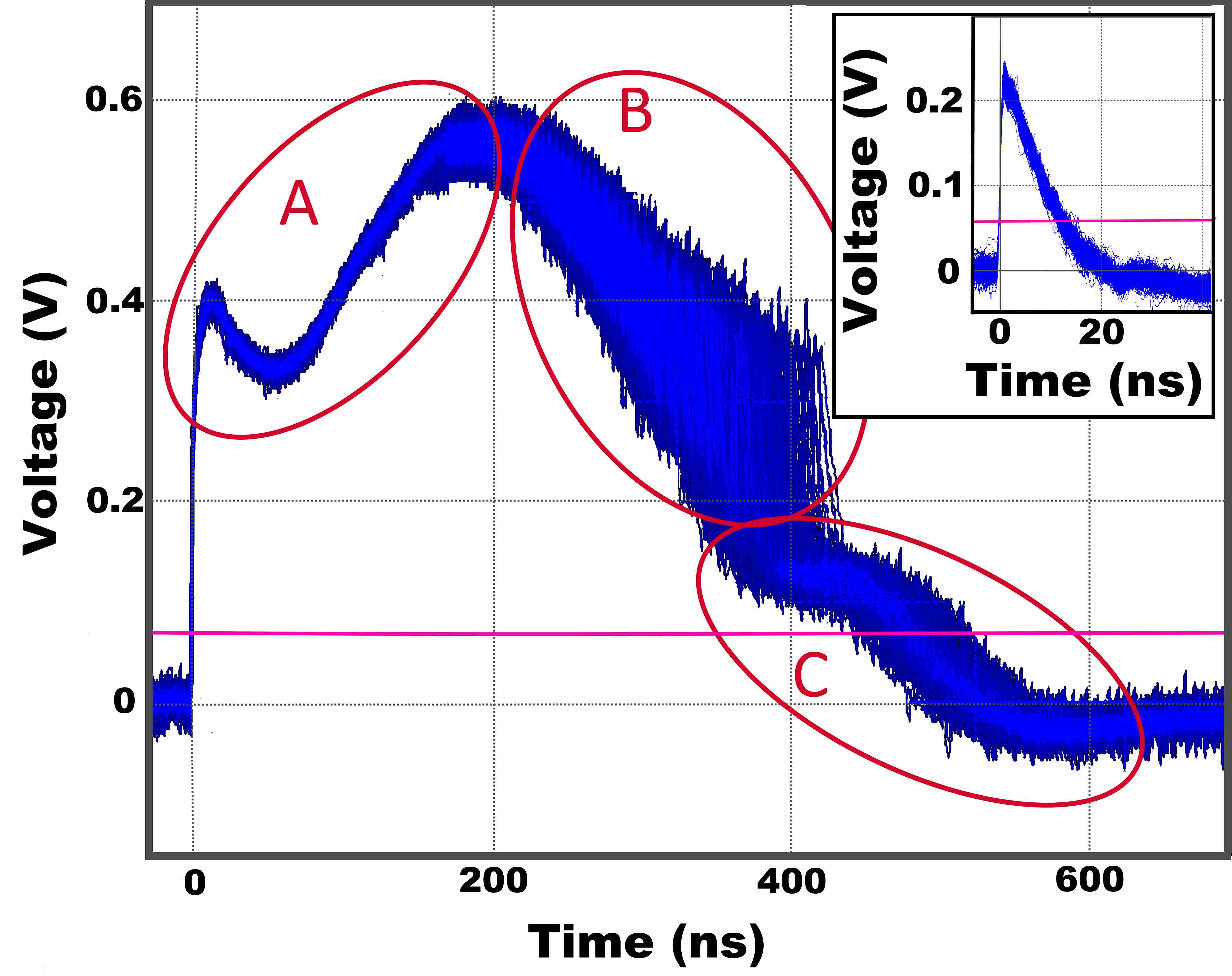}
\caption{\label{fig:pulses} Response of SNSPD to light. The oscillogram shows analog signal after the amplifiers under a bright peak power ($1.5~\milli\watt$), $100~\nano\second$ long optical pulse. The inset shows a single-photon response. The horizontal line denotes comparator threshold level. Region~A corresponds to the heating of the superconducting strip, region~B to the metastable state, and region~C to slow recovery of the superconducting state. The traces shown are a superposition of many individual oscillograms, taken in a persistence display mode.}
\end{figure}

To study how the auto-reset system reacts to the blinding attack, we have observed the SNSPD response to optical pulses of different duration $\tau$, peak power $P_p$, and repetition rate $f$. We varied $f$ in the range $1$ to $10^7~\hertz$, $P_p$ from $10^{-11}$ to $2\times10^{-4}~\watt$, and $\tau$ from $1~\nano\second$ to $1/f$ (i.e.,\ from $1~\nano\second$ wide pulses to c.w.\ illumination). \Cref{fig:pulses} shows a response to a bright $100~\nano\second$ optical pulse that does not cause the latching. A single-photon response is shown for comparison in the inset. Note that the electrical response pulse is substantially longer than the bright optical pulse, and its second half contains high-frequency peaks. The higher $P_p$ is, the longer the response pulse becomes.

This response to the bright optical pulse can be explained by the evolution of the detector's operating point (see \cref{fig:IV_curve}). Initially, the detector is at point~A. When the optical pulse hits it, it passes to point~B. This happens regardless of the optical power, as long as photons trigger the formation of normally-conducting nanowire. The transition from A to B proceeds along a $50~\ohm$ load line, because the bias source impedance at higher frequencies above $10~\mega\hertz$ (which correspond to characteristic frequencies of the single-photon response) is determined by the impedance of the signal back-end, consisting of a $50~\ohm$ coaxial cable and a matched amplifier input impedance.

In the case of a small (single-photon) optical pulse, the detector returns to point~A after a short time determined by the kinetic inductance of the nanowire. However, in the case of a high absorbed power, the temperature of the electron subsystem of the superconductor increases substantially above $T_{c}$ and takes additional time to cool down. The state of the detector begins to shift towards point~C. This is because the output impedance of the bias source at a constant current is very high (i.e.,\ it is a current source at low frequencies). Therefore, over time the source impedance gradually increases from $50~\ohm$ towards infinity, which leads to a gradual decrease of the slope of the load curve and, accordingly, a shift of the operating point from B to~C. This slow transition manifests as a smooth increase in the amplitude of the response in region~A in \cref{fig:pulses}. An additional factor is Joule heating in the normally-conducting portion of the nanowire. As the operating point shifts from B to C, the Joule heat generated continuously increases. If it exceeds the thermal power removed by dissipation from the normally conducting region, the process becomes irreversible and the detector reaches the latched state (point~C), where it will remain until it is externally reset to the superconducting state. The auto-reset system registers a control attempt only after the detector has latched after a considerable time ($\sim 10~\milli\second$).

\begin{figure}
\centering
\includegraphics[width=86mm]{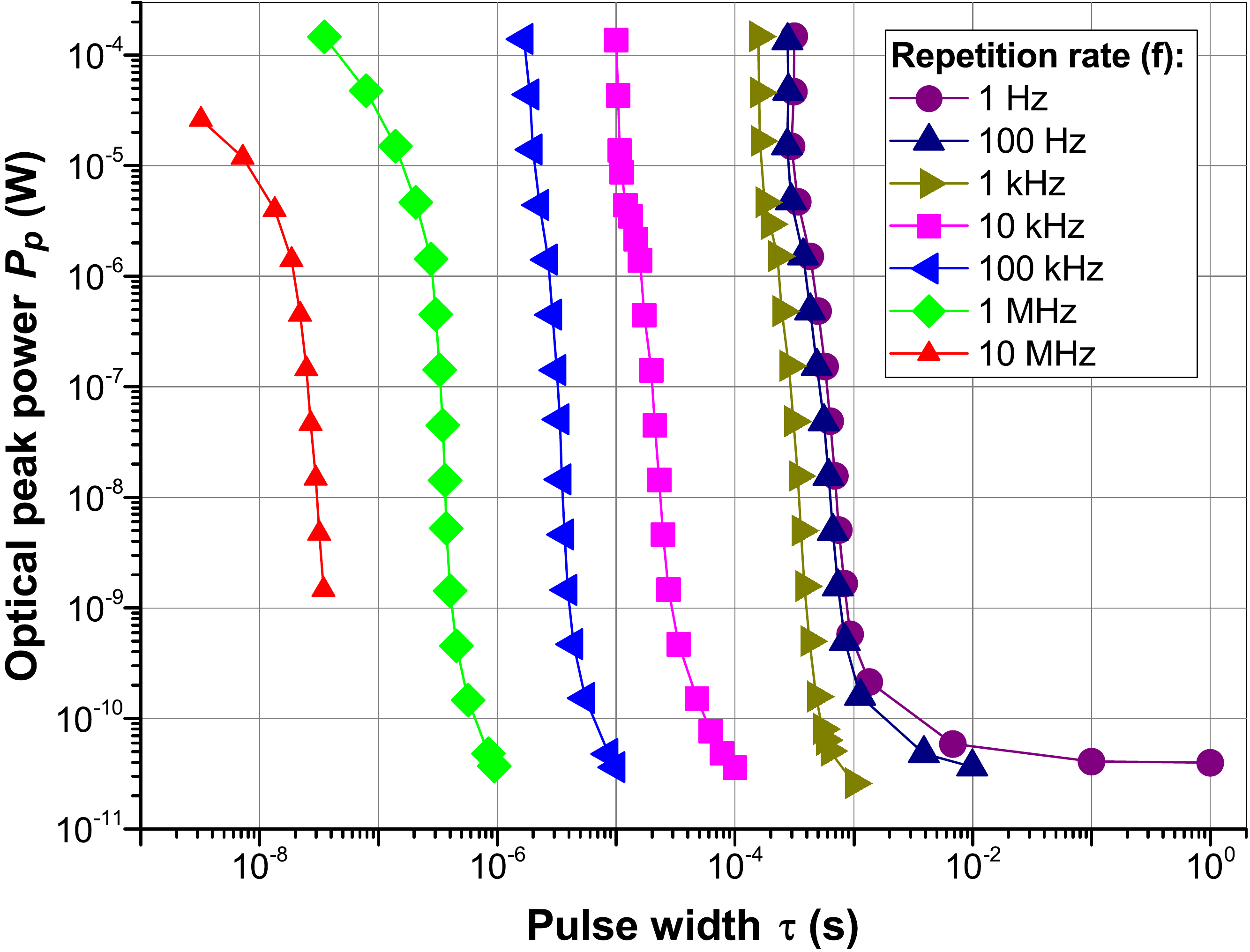}
\caption{\label{fig:Latching_threshold} Minimum optical pulse width that leads to latching at different pulse repetition rates and peak powers. When the pulse width is under the plotted value, the latching is not observed and the auto-reset system does not trigger. This regime may be exploited by Eve.}
\end{figure}

If the absorbed optical energy is insufficient for latching the detector, the temperature of its electronic subsystem decreases and, at some point, the heated nanowire goes from the normal state to a metastable one in which both the superconducting phase and the normal phase of the superconductor co-exist along the length of the nanowire \cite{liu2012}. In this mode, radio frequency generation is observed, which appears in \cref{fig:pulses} as sharp peaks in region~B. Next, the nanowire continues to cool and relatively slowly enters the superconducting state, which corresponds to region C of \cref{fig:pulses}. The shape of the electrical pulse strongly depends on the specific implementation of the bias and amplifier circuits.

For studying the latching effect, we have fixed $P_p$ and $f$. We then slowly increase $\tau$ until the detector starts latching and the auto-reset system triggers. \Cref{fig:Latching_threshold} shows this latching boundary for different $f$ and $P_p$. The last point with the maximum $\tau$ on each curve is c.w.\ power incident on the detector (except the $10~\mega\hertz$ curve where we did not reach this point because of a limitation of our PG). In our detector, latching appears at $\sim 60~\pico\watt$ c.w.\ optical power.

As this data shows, our auto-reset system is not always able to register a blinding pulse. If Eve uses optical pulses of power $P < P_p (\tau ,f)$, then her presence may not be noticed.

\begin{figure}
\centering
\includegraphics{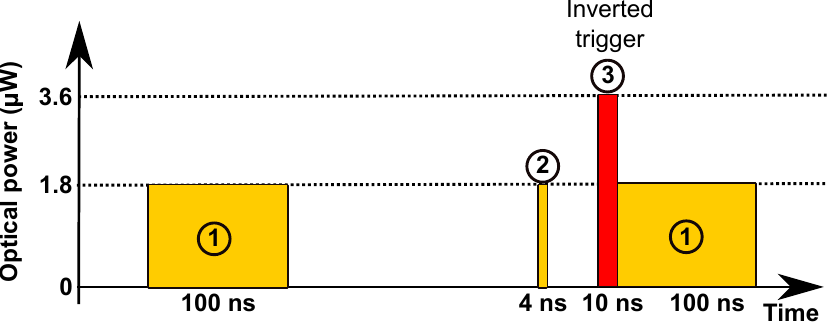}
\caption{\label{fig:Optical-burst} Faked state consisting of a sequence of pulses. The pulses \circled{1} blind the SNSPD. \circled{2} sharpens the falling edge of the nanowire recovery response. \circled{3} is an `inverted trigger' whose absence induces a click in the detector, whereas its presence keeps it blinded. In our experiment, \circled{1} and \circled{2} are generated by LD1 and \circled{3} by LD2.}
\end{figure}

\subsection{Faked-state attack on QKD}

The next step towards the faked-state attack is to demonstrate that Eve may induce clicks conditioned on Bob's basis choice \cite{makarov2009,lydersen2010a}. Let's assume the QKD system uses a Bennett-Brassard 1984 (BB84) protocol \cite{bennett1984} with polarisation encoding, an active basis choice and two detectors at Bob \cite{bennett1992b,wang2016}. For it, we have devised a pulse sequence shown in \cref{fig:Optical-burst}. First, Eve sends a bright optical pulse \circled{1} about $100~\nano\second$ long, with polarisation orthogonal to both bases used in the system, which ensures it is always equally split between Bob's detectors. Its power should be sufficient to temporarily blind Bob's detectors but not lead to their latching. Then, when the detector response is decreasing, another shorter $\sim 4~\nano\second$ pulse \circled{2} is applied. The latter pulse induces a fast recovery of the superconducting state of the nanowire. We have noticed that restoring the operating point of the SNSPD takes much longer after application of high optical power, compared to a small single-photon-like signal (see \cref{fig:pulses}). This can be explained by a constant production of Joule heat in the normal region of the nanowire. If a small optical pulse \circled{2} is sent to the slowly recovering detector, its operating point on the $I$--$V$ curve instantly shifts along the $50~\ohm$ load line, which leads to a decrease in the Joule heat produced and a faster recovery. The energy of \circled{2} has to be chosen carefully, owing to a delicate balance between the decrease of the Joule heat and the additional absorbed optical power.

\begin{figure}
\centering
\includegraphics[width=133mm]{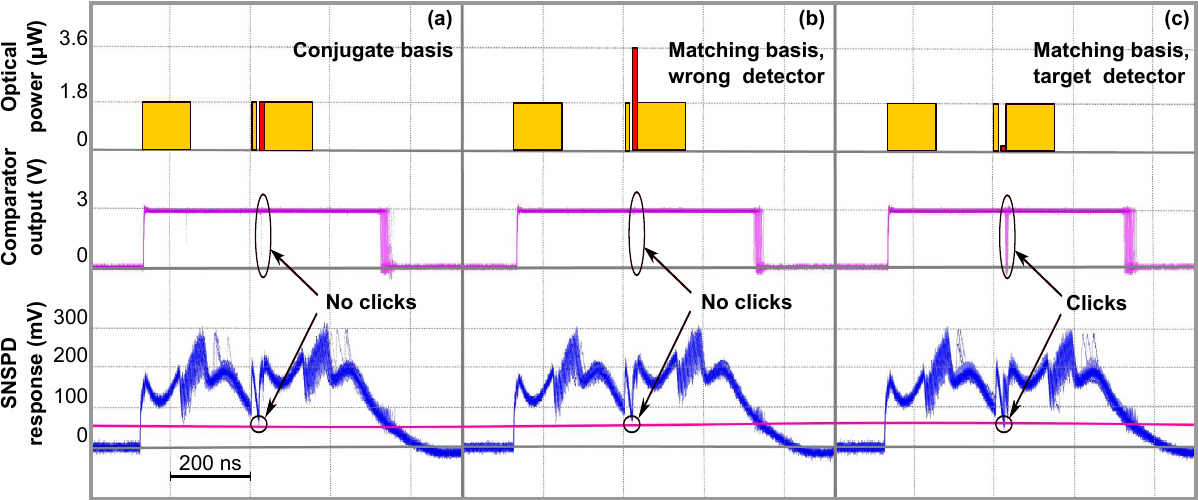}
\caption{\label{fig:Blind-Attack} Proof-of-principle of a faked-state blinding attack. The oscillograms show the response of a single SNSPD illuminated with pulse sequences that simulate those Bob's detectors would receive during the attack. (a)~Bob's basis choice does not match Eve's. Both Bob's detectors receive half the power of \circled{3}, and neither of them clicks. (b)~~Bob's basis choice matches Eve's. One of his detectors receives almost all of \circled{3} and does not click. (c)~The other Bob's detector receives a small fraction of the power of \circled{3} (here we have simulated $23~\deci\bel$ extinction ratio), and produces a click with a significant probability. The traces shown are a superposition of many individual oscillograms, taken in a persistence display mode.}
\end{figure}

Just before the detector response is about to cross below the comparator threshold, an `inverted trigger' $10~\nano\second$ pulse \circled{3} is sent, its polarisation being orthogonal to Eve's measurement result of Alice's photon. Immediately after it, a new blinding pulse \circled{1} follows, starting the next faked-state cycle. If Bob's measurement basis differs from that of Eve, \circled{3} splits evenly between Bob's detectors, blinds them and prevents them from clicking. We have simulated this regime with our single SNSPD, see \cref{fig:Blind-Attack}(a). If Bob's measurement basis matches Alice's, \circled{3} impinges on Bob's detector orthogonal to Eve's measurement, also preventing it from clicking [\cref{fig:Blind-Attack}(b)]. However the other Bob's detector receives little power determined by an extinction ratio of Bob's receiver (we have simulated $23~\deci\bel$ in our experiment), continues to recover, crosses the comparator threshold, and clicks [\cref{fig:Blind-Attack}(c)]. Thus the basis and bit value of Bob's registered clicks always matches those of Eve, making the attack possible \cite{makarov2009,lydersen2010a,gerhardt2011}.

Detector control demonstrated in our test is far from ideal. The click probability in the mismatched basis [\cref{fig:Blind-Attack}(a)] is non-zero, and in the matched basis is less than~1. However the imperfect click probabilities might not prevent the attack \Chan{\cite{lydersen2011b,qian2018}}. The controlled click has a significant jitter, apparent in the oscilloscope trace in \cref{fig:Blind-Attack}(c). If this jitter is larger than Bob's bit slot in the QKD system under attack, this may be a problem for Eve. The initial blinding pulse \circled{1} causes a simultaneous click in Bob's detectors. If its timing is accepted as a valid detection, a correctly implemented QKD system would assign a random bit value to it, which would increase the QBER under attack. To mitigate this, Eve would need to chain multiple controlled clicks under a single long blinding sequence of faked states. Our faked state structure should allow this (note that the responses to the first and second pulse \circled{1} in \cref{fig:Blind-Attack} seem to be highly similar), however we have not verified longer sequences experimentally. Finally the faked-state pulse sequence will need to be modified for implementations other than polarisation-encoding QKD. These are just a few technical challenges Eve would face when implementing a full attack.

\section{Suggestions for improving countermeasures}
\label{sec:discussion}

The objective of this study has been evaluating our security countermeasure. The detector control demonstrated is imperfect and presents Eve challenges in building her attack. Nevertheless we conclude that our countermeasure in its present form has profoundly failed the quality test. Indeed, how could the security of QKD be guaranteed by a proof if the detector is still blindable for a significant fraction of its operation time, and also controllable to a significant extent? These detector properties would be virtually impossible to account for in a security proof, necessitating a further development of the countermeasure \cite{maroy2017}.

One easy modification of the present countermeasure should be considered. In the normal photon counting operation, the comparator output is in the high logic state for less than $20~\nano\second$ (see inset in \cref{fig:pulses}). However under our demonstrated attack, it remains high for a much longer continuous time (see \cref{fig:Blind-Attack}). This condition should be detected and become an additional input to the QKD system, indicating to it that a part of the key may have been compromised and should be discarded.

\section{Conclusion}
\label{sec:conclusion}

We have developed an active anti-latching system for SNSPDs. The latching effect, previously a security vulnerability for QKD, is easily managed using this system. However, we have also found that we can still manipulate the single-photon detector to a large extent, through application of pulsed bright light. Since our goal is to create ultimately an attack-proof detector, the partial controllability is not acceptable. To further develop our countermeasure, it is necessary to monitor the comparator pulse shape and implement post-processing key extraction that correctly processes occasional alarm activations.

\section*{Funding}
This work was supported by Industry Canada, NSERC, CFI, Ontario MRI, Russian Science Foundation (project 17-72-30036, which funded the experiment), and the Ministry of Education and Science of Russia (program NTI center for quantum communications).

\section*{Acknowledgments}
We thank M.~Bourgon and J.~G.~Lim for technical help in preparing the experimental setup. 

\section*{Author contributions} 
M.E.\ conducted the experiment. R.O.\ and M.E.\ analysed the data. R.O.,\ M.E., and V.M.\ wrote the Article. V.M.\ and G.G.\ supervised the study.

\section*{Disclosures}
M.E.,\ R.O.\ and G.G.\ are interested in commercialisation of the results.
 
\bibliography{library}

\end{document}